\documentclass[11pt]{article}
\usepackage{amsfonts}
\usepackage{mathrsfs}
\usepackage{latexsym,amsmath}
\usepackage{amssymb,array}
\usepackage[sort&compress,round,comma]{natbib}
\makeatletter \oddsidemargin 0in \evensidemargin 0in \textwidth 16cm
\RequirePackage[dvips]{graphicx} \textheight 20cm
\begin{document}
\newcommand{\om}{\omega}
\newtheorem{thm}{Theorem}[section]
\newtheorem{remark}{Remark}[section]
\newtheorem{counterexample}{Counterexample}[section]
\newtheorem{note}{Note}[section]
\newtheorem{definition}{Definition}[section]
\newtheorem{example}{Example}[section]
\newtheorem{lem}{Lemma}[section]
\numberwithin{equation}{section}
\date{\today}
\title{ Some Reliability Properties of Transformed-Transformer Family of Distributions}
\author{}
\author{Nil Kamal Hazra$^1$, Pradip Kundu$^2$ and Asok K. Nanda$^2$\footnote
    {Corresponding author, e-mail:
    asok.k.nanda@gmail.com}\\
    $^1$ Department of Mathematical Statistics and Actuarial Science
\\ University of the Free State, 339 Bloemfontein 9300, South Africa\\
$^2$ Department of Mathematics and Statistics, IISER Kolkata\\
Mohanpur 741246, India\\} \maketitle
\begin{abstract}
The Transformed-Transformer family of distributions are the
resulting family of distributions as transformed from a random
variable $T$ through another transformer random variable $X$ using a
weight function $\omega$ of the cumulative distribution function of
$X$. In this paper, we study different stochastic ageing properties,
as well as different stochastic orderings of this family of
distributions. We discuss the results with several well known
distributions.
\end{abstract}
\section{Introduction}
Different probability distributions have been developed in the
literature by different researcher to give more flexibility in
modelling and data analysis. Sometimes these developments are
data-driven, and sometimes they are theory-driven. Most of the
methods developed to generate new family of distributions generally
fall into two categories - combining existing distributions into new
distributions and adding new parameters to an existing distribution.
Although, in general, more number of parameters in a probability
distribution gives rise to more flexibility in modelling, after
sometime the increase in the number of parameters leads to quite
marginal improvement as far as flexibility in the analysis is
concerned. It is observed that although at least three parameters
are required for a probability distribution to have some practical
usefulness, the optimum number of parameters is four in the sense
that the increase in the number of parameters beyond four does not
give any substantial improvement (Johnson et al.\citealp{joh};
Alzaatreh et al.\citealp{alz2}). System of probability distributions
developed by Karl Pearson (Elderton and Johnson\citealp{eld}), and
Burr family of distributions developed by Burr\cite{bur} are two
such well known family of distributions. Though generalized lamda
distribution (Ramberg and Schmeiser\citealp{ram1}; Ramberg et
al.\citealp{ram2}), skew-normal family of distributions
(Azzalini\citealp{azz}), beta-generated distributions (Akinsete et
al.\citealp{aki}; Alshawarbeh et al.\citealp{als}; Barreto-Souza et
al.\citealp{bar1,bar2}; Cordeiro et al.\citealp{cor2}; Eugene et
al.\citealp{eug}; Famoye et al.\citealp{fam}; Kong et
al.\citealp{kon}; Lee et al.\citealp{lee}; Nadarajah and
Kotz\citealp{nad}), Kumaraswamy-generated distribution
(Jones\citealp{jon}; Cordeiro and de Castro\citealp{cor1}) among
others are the recent addition in the family of distributions, a
more recent development is the Transformed-Transformer family of
distributions studied by Alzaatreh et al.\cite{alz2,alz1,alz3}, who
call it $T$-$X$ family of distributions. This family of
distributions is generated by transformation from a random variable
$T$ through another random variable $X$ using weight function
$\omega$ of the cumulative distribution function of $X$. This is the
reason to call it Transformed-Transformer family of distributions.
Different choices of $T$, $X$ and $\omega$ lead to different
families of distributions. Below we give a brief discussion on this.
\par Let $T$ be an absolutely continuous random variable with
support $[a,b]$, where $-\infty< a< b<\infty$ and let $X$ be another
random variable with support $[c,d]$, where $-\infty< c<d< \infty$.
Further, let $\om_1:[0,1)\rightarrow[a,b]$ be a continuous function
such that
\begin{enumerate}
 \item [$(i)$] $\om_1(\cdot)$ is differentiable and monotonically increasing;
 \item [$(ii)$] $\om_1(0)=a$ and $\lim\limits_{x\rightarrow 1-}\om_1(x)=b$.
\end{enumerate}
For a random variable $Z$, we denote the probability density
function (p.d.f.) of $Z$ by $f_{Z}$ with cumulative distributive
function (c.d.f.) $F_{Z}$ and survival function $\bar{F}_{Z}$. Then
the c.d.f. $F(\cdot)$ of the Transformed-Transformer family of
distributions is defined, for $x\in[c,d]$, as
\begin{eqnarray}
  F(x)&=&\int\limits_a^{\om_1(F_X(x))} f_T(u)du\nonumber
 \\\nonumber
 \\&=& F_T[\om_1(F_X(x))].\label{eq1}
\end{eqnarray}
Let the corresponding random variable be denoted by $R$. The
reliability function of $R$ is given by
\begin{eqnarray}
  \bar F(x)&=&\int\limits_{\om_1(F_X(x))}^b f_T(u)du\nonumber
 \\\nonumber
 \\&=& \bar F_T[\om_1(F_X(x))]\nonumber
 \\&=& \bar F_T[\om_2(\bar F_X(x))],\label{eq01}
\end{eqnarray}
where $\om_2(x)=\om_1(1-x)$.\par Alzaatreh et al.\cite{alz2}
obtained different distributions for different choice of the
distributions of $T$ and $X$ based on different weight function
$\omega(\cdot)$. It is interesting to note (Alzaatreh et
al.\citealp{alz3}) that for any random variable with support in
$(a,b)$, $\omega(\cdot)$ can be taken as the quantile function of
distribution of that random variable.\par One may notice that a
large number of distributions appear as special cases of the
Transformed-Transformer family of distributions, viz. beta-generated
family of distributions, exponential, gamma, Weibull, Lomax,
Rayleigh, generalized gamma, exponentiated-Weibull, gamma-Pareto,
exponentiated-exponential and many more distributions. Different
properties of distributions have been separately studied in the
literature by different researchers. The Transformed-Transformer
family of distributions, being quite general, the properties studied
here will hold true for all those distributions which are members of
this family (as particular cases).\par In this article we study
different reliability properties of the Transformed-Transformer
family of distributions taking a general weight function
$\omega_{1}$ or $\omega_{2}$. The paper is organized as follows.
Different stochastic ageing properties of this family of
distributions are studied in Section 2. Section 3 deals with
different stochastic orders. We give a brief conclusion in Section
4.\par For any function $g$, we write $g'(x)$ to denote the first
order derivative of $g$ with respect to $x$. The word increasing and
decreasing are not used in strict sense.
\section{Stochastic Ageing properties}
In this section we throw some light on how different ageing
properties of $T$ and $X$ are transmitted to the random variable $R$
through the weight function $\omega_{2}$ (or $\omega_{1}$).\par
Below we see that, under certain condition on $\omega_{2}$, the IFR
(increasing failure rate) property of $T$ and $X$ is transmitted to
the random variable $R$. It is to be mentioned here that a random
variable $Z$ is said to be IFR if the failure rate function defined
as the ratio of the p.d.f. to its survival function, is increasing,
i.e., if $r_{Z}(t)=f_{Z}(t)/\bar{F}_{Z}(t)$ is increasing in $t$.
\begin{thm}\label{th2.1}
 Let $x\omega'_2(x)$ be increasing in $x\in(0,1]$. If $X$ and $T$ are IFR then $R$ is IFR.
\end{thm}
{\bf Proof:} From (\ref{eq01}), the density function $f(\cdot)$ of
$R$ is given by
\begin{eqnarray*}
 f(x)=f_T[\om_2(\bar F_X(x))]\frac{d}{dx}\left(\om_2(\bar F_X(x))\right),
\end{eqnarray*}
which gives the corresponding failure rate as
\begin{eqnarray*}
 r(x)=-r_T[\om_2(\bar F_X(x))]r_X(x)\bar F_X(x)\om_2'(\bar F_X(x)).\label{eq11}
\end{eqnarray*}
Since $T$ is IFR, we have that $r_T[\om_2(\bar F_X(x))]\;\text{is
increasing in }x\in[c,d].$ Thus, to prove that $R$ is IFR, it
suffices to show that
\begin{eqnarray}
 r_X(x)\bar F_X(x)\left(-\om_2'(\bar F_X(x))\right)\;\text{is increasing in }\;x\in[c,d].\label{eq13}
\end{eqnarray}
Since $X$ is IFR, we have that $r_X(x)\;\text{is increasing in }x
\in[c,d].$ Thus, (\ref{eq13}) holds if
\begin{eqnarray*}
 \bar F_X(x)\om_2'(\bar F_X(x))\;\text{is decreasing in }x\in[c,d],
\end{eqnarray*}
or equivalently,
\begin{eqnarray*}
x\omega'_2(x)\;\text{is increasing in }x\in(0,1].
\end{eqnarray*}
Hence, the result follows.
\begin{remark}\normalfont
 It is easy to see that $x\omega'_2(x)$ is increasing in $x\in(0,1]$ if, and only if, $(1-x)\omega'_1(x)$ is increasing in $x\in[0,1)$.
\end{remark}
\begin{remark}\normalfont
 It is to be noted that (i) $\omega_2(x)=-\ln x$, (ii)
$\omega_2(x)=(1-x)^\alpha/x^{\alpha-1}$, for $\alpha\geq 1$, (iii)
$\omega_2(x)=(1-x)/x$, (iv) $\omega_2(x)=(1-x^2)/x$, (v)
$\omega_2(x)=(1-x)^2/(1-(1-x)^2)$ satisfy the condition given in
Theorem \ref{th2.1}.
\end{remark}
\par Following counterexample shows that the condition
`$x\omega'_2(x)$ is increasing in $x\in(0,1]$' in the above theorem
cannot be dropped.
\begin{counterexample}\normalfont
Take $\omega_2(x)=(1-x^3)/\sqrt{x}$, $x\in(0,1]$ so that
$x\omega'_2(x)$ is neither increasing nor decreasing in $x\in(0,1]$.
Let $X$ follow Weibull distribution with distribution function
$F_{X}(x)=1-e^{-(x/\beta)^k}$, $x\geq 0$, $\beta>0$, $k\geq 1$, and
let $T$ follow exponential distribution with distribution function
$F_{T}(x)=1-e^{-\gamma x}$, $x\geq 0$, $\gamma>0$. Then the failure
rate function of the random variable $R$ is given by
$$r(x)=\frac{\gamma k x^{k-1} (5e^{-3(x/\beta)^k +1})}{2\beta^k
\sqrt{e^{-(x/\beta)^k}}}.$$ Now, it can be easily verified that
 $r(x)$ is neither increasing nor decreasing in $x\geq
0$, for $\gamma=2$, $\beta=0.4$, $k=2$. This shows that $R$ is not
IFR.$\hfill\Box$
\end{counterexample}
\par Using Theorem \ref{th2.1} and different $\omega_2(x)$ as given in
Remark 2.2, one can generate large number of IFR distributions
taking different $T$ and $X$ having IFR property. Below we discuss
some distributions where IFR property of gamma distribution is
transmitted to $R$.
\begin{example}\normalfont
Take $\omega_{1}(x)=-\ln(1-x)$ for $x\in[0,1)$, i.e.
$\omega_{2}(x)=-\ln~x$ for $x\in(0,1]$. Let $T$ follow gamma
distribution with p.d.f. given by
\begin{equation}\label{gamma}f_{T}(t)=\frac{1}{\Gamma (\alpha)
\lambda^\alpha}t^{\alpha-1}e^{-t/\lambda},~t>0,~\alpha,
\lambda>0.\end{equation} Note that $T$ is IFR if $\alpha\geq 1$.
From (\ref{eq1}), we have the p.d.f. of $R$ as
\begin{eqnarray} \nonumber f(x)&=&\frac{f_{X}(x)}{1-F_{X}(x)}
f_{T}[-\ln(1-F_{X}(x))]\\&=&\label{gamma-x}\frac{1}{\Gamma (\alpha)
\lambda^\alpha}f_{X}(x)[-\ln(1-F_{X}(x))]^{\alpha-1}(1-F_{X}(x))^{\frac{1}{\lambda}-1}.
\end{eqnarray}
For different distributions of $X$, using Theorem \ref{th2.1}, we
get the IFR property of the gamma-$X$ family of distributions with
p.d.f. given in (\ref{gamma-x}), as presented in Table 1.\par We can
generate more gamma-$X$ family of distributions with IFR property by
taking different $\omega_2(x)$. For instance, taking $T$ to be a
gamma random variable with p.d.f. as given in (\ref{gamma}) with
$\alpha\geq 1$, and $X$, a Weibull distribution with $k\geq 1$ as
given in Table 1, we present different gamma-Weibull family of
distributions satisfying IFR property for different $\omega_2(x)$ in
$\text{Table 2}$. One can similarly generate gamma-exponential,
gamma-half normal, gamma-Gompertz, gamma-Makeham, gamma-Rayleigh
family of distributions each of which is IFR. The newly generated
distributions can be seen as a generalization of the many well-known
distributions. For example, consider the gamma-Weibull distribution
generated by using $\omega_2(x)=-\ln x$ with p.d.f. given in Table
2. Then putting $\lambda=1$ we get the generalized gamma
distribution discussed by Stacy\cite{stacy} (also see Khodabin and
Ahmadabadi\citealp{khob}); for $\beta=k=1$ or $\lambda=k=1$, we get
gamma distribution; for $k=2$, $\alpha=1/2$, $\lambda=1$, setting
$\beta^2=2 \sigma^2$ we get half-normal distribution; for $k=2$,
$\alpha=\lambda=1$, setting $\beta^2=2 \sigma^2$ we get Rayliegh
distribution.$\hfill\Box$
\begin{table}[h]
\begin{center}{Table 1: Ageing property of gamma-$X$ family of
distributions\\
\begin{tabular}{|ll|}
\hline  \multicolumn{1}{|l}{Distribution of $X$} &
\multicolumn{1}{c|}{Ageing property of $R$ }
\\\hline Exponential & IFR for $\alpha\geq 1$
 \\$F_{X}(x)=1-e^{-\beta x}$, $x\geq 0$, $\beta>0$  &
 \\\hline
 Weibull  &
IFR for $\alpha\geq 1$, $k\geq 1$\\$F_{X}(x)=1-e^{-(x/\beta)^k}$,
$x\geq 0$, $k, \beta>0$ &
 \\\hline $\text{Half Normal}^\star$
& IFR for $\alpha\geq 1$
 \\
 $F_{X}(x)=erf(\frac{x}{\sigma \sqrt2}),$ $x>0$, $\sigma>0$
&  \\\hline Gompertz & IFR for $\alpha\geq 1$, $c\geq1$
\\
$F_{X}(x)=1-e^{-B(c^x-1)/\ln c}$, $x,c\geq 0$, $B>0$ & \\\hline
Makeham & IFR for $\alpha\geq
1$\\
$F_{X}(x)=1-e^{[-\beta x+(\gamma/\eta)(e^{\eta x}-1)]}$, &\\$x\geq
0$, $\beta, \gamma, \eta>0$ &
\\\hline
Rayleigh & IFR for $\alpha\geq
1$\\
$F_{X}(x)=1-e^{-x^{2}/2\sigma^{2}}$, $x\geq 0$ &
\\\hline
\end{tabular} }
\\$\star$ $erf(\cdot)$ is the error function defined as
$erf(x)=\frac{2}{\sqrt{\pi}}\int_{0}^x e^{-t^{2}}dt$.
\end{center}
\end{table}
\begin{table}[h]
\begin{center}{Table 2: Gamma-Weibull family of
distributions having IFR property\\
\begin{tabular}{|ll|}
\hline  \multicolumn{1}{|c}{$\omega_2(x)$} &
\multicolumn{1}{l|}{p.d.f. of gamma-Weibull family of distributions
with $\alpha\geq 1$, $k\geq 1$}
\\\hline $-\ln x$  &
$\frac{k}{\Gamma (\alpha) \lambda^\alpha \beta}
\left(\frac{x}{\beta}\right)^{k
\alpha-1}e^{-\frac{1}{\lambda}\left(\frac{x}{\beta}\right)^k}$
\\ $(1-x)/x$ & $\frac{k}{\Gamma (\alpha)
\lambda^\alpha \beta} e^{\left(\frac{x}{\beta}\right)^k}
\left(\frac{x}{\beta}\right)^{k-1}
\left(e^{\left(\frac{x}{\beta}\right)^k}-1\right)^{\alpha-1}
e^{-\frac{1}{\lambda}\left(e^{\left(\frac{x}{\beta}\right)^k}-1\right)}$
 \\ $(1-x)^2/x$ & $\frac{k\left(2-e^{-\left(\frac{x}{\beta}\right)^k}\right)}{\Gamma (\alpha)
\lambda^\alpha \beta} \left(\frac{x}{\beta}\right)^{k-1}
\left(\frac{1-e^{-\left(\frac{x}{\beta}\right)^k}}{\sqrt{e^{-\left(\frac{x}{\beta}\right)^k}}}\right)^{2(\alpha-1)}
e^{-\frac{1}{\lambda}\left(\frac{1-e^{-\left(\frac{x}{\beta}\right)^k}}{\sqrt{e^{-\left(\frac{x}{\beta}\right)^k}}}\right)^{2}}$\\
$(1-x^2)/x$ & $\frac{k
\left(e^{\left(\frac{x}{\beta}\right)^k}+e^{-\left(\frac{x}{\beta}\right)^k}\right)}{\Gamma
(\alpha) \lambda^\alpha
\beta}\left(\frac{x}{\beta}\right)^{k-1}\left(e^{\left(\frac{x}{\beta}\right)^k}-e^{-\left(\frac{x}{\beta}\right)^k}\right)^{\alpha-1}e^{-\frac{1}{\lambda}\left(e^{\left(\frac{x}{\beta}\right)^k}-e^{-\left(\frac{x}{\beta}\right)^k}\right)}
$\\\hline
\end{tabular} }
\end{center}
\end{table}\label{tab2}
\end{example}
\par Below we see how NBU (new better than used) property of $T$ and
$X$ is transmitted to the random variable $R$. It is to be noted
that a random variable $Z$ is said to be NBU if
$\bar{F}_{Z}(x+t)\leq \bar{F}_{Z}(x) \bar{F}_{Z}(t)$, for all $x$,
$t$.
\begin{thm}\label{th2.2}
 Let $\omega_2(xy)\geq \om_2(x)+\om_2(y)$ for all $x,y\in(0,1]$. If $X$ and $T$ are NBU then $R$ is NBU.
\end{thm}
{\bf Proof: }Since $X$ is NBU, we have, for all $x,t\in[c,d]$,
\begin{eqnarray*}
 \om_2(\bar F_X(x+t))&\geq& \om_2(\bar F_X(x))+\om_2(\bar F_X(t)),
\end{eqnarray*}
which, on using the fact that T is NBU, gives
\begin{eqnarray*}
 \bar F_T\left(\om_2(\bar F_X(x+t))\right)&\leq& \bar F_T\left(\om_2(\bar F_X(x))\right)\bar F_T\left(\om_2(\bar
 F_X(t))\right).
\end{eqnarray*}
Hence, the result follows.
\begin{remark}\normalfont
 The condition of Theorem \ref{th2.2} is satisfied by (i) $\omega_2(x)=-\ln x$, (ii) $\omega_2(x)=(1-x)/x$, (iii)
$\omega_2(x)=(1-x)^2/x$.
\end{remark}
\begin{remark}\normalfont
Since IFR implies NBU, the distribution of $R$ given in Table 1 and
Table 2 are NBU. Consider $X$ having the following distribution.
$$F_{X}(x)=1-e^{-h(x)},~x\geq 0,$$ where $$h(x)=\left\{
                                                                    \begin{array}{ll}
                                                                      \sin^2 x, & \hbox{if $x\in(0,\frac{\pi}{2}]$;} \\
                                                                      \frac{1}{2}\pi(x-\frac{1}{2}\pi)+1, & \hbox{if $x\in(\frac{\pi}{2},\infty)$.}
                                                                    \end{array}
                                                                  \right.$$
Then $X$ is NBU (but not IFR). So, if $T$ is NBU, then the resulting
family of distributions is NBU, for any $\omega_2(x)$ given in
Remark 2.3. For instance, if $T$ follows gamma distribution with
density function as in (\ref{gamma}) with $\alpha\geq 1$, then, for
$\omega_2(x)=-\ln x$, we have a NBU distribution with p.d.f.
$f_{X}(x)=\frac{1}{\Gamma (\alpha) \lambda^\alpha} (h(x))^{\alpha-1}
h'(x) e^{-\frac{1}{\lambda}h(x)}$.$\hfill\Box$
\end{remark}
\par The following counterexample shows that the condition
`$\omega_2(xy)\geq \om_2(x)+\om_2(y)$ for all $x,y\in(0,1]$' in
Theorem \ref{th2.2} cannot be dropped.
\begin{counterexample}\normalfont
Take $\omega_2(x)=\left(\frac{1-x^3}{\sqrt{x}}\right)^{1/2}$,
$x\in(0,1]$. It can be seen that $\omega_2(xy)\ngeq
\om_2(x)+\om_2(y)$, $\forall~x\in(0,1]$. Let $X$ follow Gompertz
distribution with distribution function $F_{X}(x)=1-e^{-B(c^x-1)/\ln
c}$, $x\geq 0$, $B>0$, $c\geq 1$, and let $T$ follow Weibull
distribution with distribution function
$F_{T}(x)=1-e^{-(x/\beta)^k}$, $x\geq 0$, $\beta>0$, $k\geq 1$. Then
$X$ and $T$ both are NBU. For $k=2$, $\beta=1$, $B=1$, $c=2$, we
have
$$\bar{F}_{T}\left(\om_2(\bar
F_X(x))\right)=\exp\left\{-\left(\frac{1-e^{-3B(c^x-1)/\ln
c}}{\sqrt{e^{-B(c^x-1)/\ln c}}}\right)\right\}.$$ Now, for $x=0.4$
and $t=0.3$, we have $$\bar{F}_{T}\left(\om_2(\bar
F_X(x+t))\right)=0.2313256~ \text{and}~ \bar{F}_{T}\left(\om_2(\bar
F_X(x))\right)\cdot\bar{F}_{T}\left(\om_2(\bar
F_X(t))\right)=0.1844626.$$ This shows that $R$ is not NBU.
\end{counterexample}
\section{Stochastic Orderings}
Let the random variable $R_{1}$ be derived from the random variables
$T_{1}$ and $X_{1}$, and let $R_{2}$ be another random variable
derived from the random variables $T_{2}$ and $X_{2}$. In both the
cases we take the same weight function $\omega_{1}$ (or equivalently
$\omega_{2}$). In this section we study how different ordering
properties between $T_{1}$ and $T_{2}$, and those between $X_{1}$
and $X_{2}$ are transformed to the ordering properties between
$R_{1}$ and $R_{2}$.\par Below we show that if $T_{2}$ dominates
$T_{1}$, and $X_{2}$ dominates $X_{1}$ in usual stochastic order,
then $R_{2}$ dominates $R_{1}$ in usual stochastic order. It is to
be mentioned here that, for random variable $Z_{i}$ having survival
function $\bar{F}_{i}$, $i=1,2$, $Z_{2}$ is said to dominate $Z_{1}$
in stochastic order, written as $Z_{1}\leq_{st} Z_{2}$, if
$\bar{F}_{1}(t)\leq \bar{F}_{2}(t)$ for all $t$.
\begin{thm}\label{th3.1}
 If $T_1\leq_{st}T_2$ and $X_1\leq_{st} X_2$, then $R_1\leq_{st} R_2$.
\end{thm}
{\bf Proof:} $X_1\leq_{st} X_2$ gives that, for all $x\in[c,d]$,
$\om_2(\bar F_{X_1}(x))\geq \om_2(\bar F_{X_2}(x)),$ which further
gives that
\begin{eqnarray*}
 \bar F_{T_1}[\om_2(\bar F_{X_1}(x))]&\leq&\;\bar F_{T_1}[\om_2(\bar F_{X_2}(x))]\nonumber
 \\&\leq&\;\bar F_{T_2}[\om_2(\bar F_{X_2}(x))],
\end{eqnarray*}
giving $R_1\leq_{st} R_2$. $\hfill\Box$\par
In the following theorem we give conditions on $T_{1}$, $T_{2}$,
$X_{1}$ and $X_{2}$, under which $R_{2}$ dominates $R_{1}$ in up
shifted hazard rate order. For two random variables $Z_{1}$ and
$Z_{2}$ having respective survival functions $\bar{F}_{1}$ and
$\bar{F}_{2}$, $Z_{2}$ is said to dominate $Z_{1}$ in up shifted
hazard rate order, written as $Z_{1}\leq_{hr \uparrow} Z_{2}$, if
$\bar{F}_{2}(x)/\bar{F}_{1}(x+t)$ is increasing in $x$ for all $t$.
\begin{thm}\label{th3.2}
 Let $x\omega'_2(x)$ be increasing in $x\in(0,1]$. Suppose that the following conditions hold:
 \begin{enumerate}
  \item [(i)] $T_{1}\leq_{hr}T_{2}$, and $T_{1}$ or $T_{2}$ is IFR;
  \item [(ii)] $X_{1}\leq_{hr\uparrow} X_{2}$.
 \end{enumerate}
Then $R_{1}\leq_{hr \uparrow} R_{2}$.
\end{thm}
{\bf Proof:} $R_1\leq_{hr \uparrow} R_2$ holds if, and only if, for
all $t\geq 0$,
\begin{eqnarray*}
 r_{T_1}[\om_2(\bar F_{X_1}(x+t))]\frac{d}{dx}\{\om_2(\bar F_{X_1}(x+t))\}\geq r_{T_2}[\om_2(\bar F_{X_2}(x))]\frac{d}{dx}\{\om_2(\bar F_{X_2}(x))\}.
\end{eqnarray*}
This holds if
\begin{eqnarray}
 r_{T_1}[\om_2(\bar F_{X_1}(x+t))]\geq r_{T_2}[\om_2(\bar F_{X_2}(x))]\label{eq2}
\end{eqnarray}
and
\begin{eqnarray}
 \frac{d}{dx}\{\om_2(\bar F_{X_1}(x+t))\}\geq \frac{d}{dx}\{\om_2(\bar F_{X_2}(x))\}.\label{eq3}
\end{eqnarray}
Since up hazard rate order is stronger than usual stochastic order,
$(ii)$ gives, for all $t\geq 0$ and for all $x\in[c,d]$,
\begin{eqnarray}
\om_2(\bar F_{X_1}(x+t))\geq \om_2(\bar F_{X_2}(x)).\label{eq6}
\end{eqnarray}
Let us consider the following two cases.
\\Case I: Let $T_1$ be IFR. Then, from (\ref{eq6}), we have, for all $t\geq 0$ and for all $x\in[c,d]$,
\begin{eqnarray}
r_{T_1}\left(\om_2(\bar F_{X_1}(x+t))\right)&\geq&
r_{T_1}\left(\om_2(\bar F_{X_2}(x))\right)\nonumber
\\&\geq& r_{T_2}\left(\om_2(\bar F_{X_2}(x))\right).\label{eq7}
\end{eqnarray}
Case II: Let $T_2$ be IFR. Then, from (\ref{eq6}), we have, for all
$t\geq 0$ and for all $x\in[c,d]$,
\begin{eqnarray}
r_{T_2}\left(\om_2(\bar F_{X_2}(x))\right)&\leq&
r_{T_2}\left(\om_2(\bar F_{X_1}(x+t))\right)\nonumber
\\&\leq& r_{T_1}\left(\om_2(\bar F_{X_1}(x+t))\right).\label{eq8}
\end{eqnarray}
Thus, from (\ref{eq7}) and (\ref{eq8}), (\ref{eq2}) holds. Note
that, (\ref{eq3}) holds if, and only if, for all $t\geq 0$ and, for
all $x\in[c,d]$,
\begin{eqnarray}
 r_{X_1}(x+t)\left[-\bar F_{X_1}(x+t)\om'_2\left(\bar F_{X_1}(x+t)\right)\right]\geq r_{X_2}(x)\left[-\bar F_{X_2}(x)\om'_2\left(\bar F_{X_2}(x)\right)\right].\label{eq9}
\end{eqnarray}
Since $X_1\leq_{hr\uparrow} X_2$, to prove (\ref{eq9}), it suffices
to show that
\begin{eqnarray}
 \bar F_{X_1}(x+t)\om'_2\left(\bar F_{X_1}(x+t)\right)\leq \bar F_{X_2}(x)\om'_2\left(\bar F_{X_2}(x)\right),
\end{eqnarray}
which holds from the fact that $x\om'_2(x)$ is increasing in
$x\in(0,1]$. Hence, the result is proved.$\hfill\Box$\par With the
following counterexample, we show that the condition
`$x\omega'_2(x)$ is increasing in $x\in(0,1]$' in Theorem
\ref{th3.2} cannot be dropped.
\begin{counterexample}\normalfont
Take $\omega_2(x)=\left(\frac{1-x^3}{\sqrt{x}}\right)^{1/2}$,
$x\in(0,1]$. It can be seen that $x\omega'_2(x)$ is not monotone.
Let $X_1$ and $X_2$ follow exponential distribution with failure
rates $\beta_{1}$ and $\beta_{2}$ respectively. Then
$X_{1}\leq_{hr\uparrow} X_{2}$ if $\beta_{1}\geq \beta_{2}$. Let
$T_1$ and $T_2$ follow Rayleigh distribution with distribution
function $F_{T_1}(x)=1-e^{-x^{2}/2\sigma_{1}^{2}}$, $x\geq 0$ and
$F_{T_2}(x)=1-e^{-x^{2}/2\sigma_{2}^{2}}$, $x\geq 0$, respectively.
Here $T_1$ and $T_2$ are IFR, and $T_{1}\leq_{hr}T_{2}$ for
$\sigma_{1}^{2}\leq \sigma_{2}^{2}$. It can be easily verified that, for $\beta_{1}=2$, $\beta_{2}=1.8$,
$\sigma_{1}=0.6$, $\sigma_{2}=0.65$ and $t=0.9$, $\bar
F_{T_2}[\om_2(\bar F_{X_2}(x))]/\bar F_{T_1}[\om_2(\bar
F_{X_1}(x+t))]$ is neither increasing nor decreasing in $x$. Thus,
$R_{2}$ does not always dominate $R_{1}$ with respect to the up
shifted hazard rate order. $\hfill\Box$
\end{counterexample}
\par The following theorem gives a result similar to that given in
Theorem~\ref{th3.2} for the hazard rate order. The proof follows in
the same line as that of Theorem~\ref{th3.2}. If $\bar{F}_{1}$ and
$\bar{F}_{2}$, the respective survival functions of random variables
$Z_{1}$ and $Z_{2}$, be such that $\bar{F}_{1}(x)/\bar{F}_{2}(x)$ is
decreasing in $x$, we say that $Z_{2}$ dominates $Z_{1}$ in hazard
rate order, and we write $Z_{1}\leq_{hr} Z_{2}$.
\begin{thm}\label{th3.3}
 Let $x\omega'_2(x)$ be increasing in $x\in(0,1]$. Suppose that the following conditions hold:
 \begin{enumerate}
  \item [(i)] $T_1\leq_{hr}T_2$, and $T_1$ or $T_2$ is IFR;
  \item [(ii)] $X_1\leq_{hr} X_2$.
 \end{enumerate}
Then $R_1\leq_{hr} R_2$.$\hfill\Box$
\end{thm}
\par For two random variables $Z_{1}$ and $Z_{2}$ having respective
reversed hazard rates $\tilde{r}_{1}$ and $\tilde{r}_{2}$ (and
respective distribution functions $F_{1}$ and $F_{2}$), $Z_{2}$ is
said to dominate $Z_{1}$ in reversed hazard rate order, written as
$Z_{1}\leq_{rhr} Z_{2}$, if $\tilde{r}_{1}(x)\leq \tilde{r}_{2}(x)$
for all $x$, or equivalently, $F_{1}(x)/F_{2}(x)$ is decreasing in
$x$. Below we give conditions under which $Z_{2}$ dominates $Z_{1}$
in reversed hazard rate order.
\begin{thm}\label{th3.4}
 Let $x\omega'_1(x)$ be decreasing in $x\in[0,1)$. Suppose that the following conditions hold:
 \begin{enumerate}
  \item [(i)] $T_1\leq_{rhr}T_2$, and $T_1$ or $T_2$ is DRHR;
  \item [(ii)] $X_1\leq_{rhr} X_2$.
 \end{enumerate}
Then $R_1\leq_{rhr} R_2$.
\end{thm}
{\bf Proof:} $R_1\leq_{rhr} R_2$ holds if, and only if,
\begin{eqnarray*}
 \tilde{r}_{T_2}[\om_1(F_{X_2}(x))]\frac{d}{dx}\{\om_1(F_{X_2}(x))\}\geq \tilde{r}_{T_1}[\om_1(F_{X_1}(x))]\frac{d}{dx}\{\om_1(F_{X_1}(x))\}.
\end{eqnarray*}
This holds if
\begin{eqnarray}
 \tilde{r}_{T_2}[\om_1(F_{X_2}(x))]\geq \tilde{r}_{T_1}[\om_1(F_{X_1}(x))]\label{eq3.41}
\end{eqnarray}
and
\begin{eqnarray}
 \frac{d}{dx}\{\om_1(F_{X_2}(x))\}\geq \frac{d}{dx}\{\om_1(F_{X_1}(x))\}.\label{eq3.42}
\end{eqnarray}
Since reversed hazard rate order is stronger than usual stochastic
order, we have, for all $x\in[c,d]$,
\begin{eqnarray}\om_1(F_{X_2}(x))\leq \om_1(F_{X_1}(x)).\label{eq3.43}
\end{eqnarray}
Let us consider the following two cases.
\\Case I: Let $T_1$ be DRHR. Then, from (\ref{eq3.43}), we have, for all $x\in[c,d]$,
\begin{eqnarray}
\tilde{r}_{T_1}\left(\om_1(F_{X_1}(x))\right)&\leq&
\tilde{r}_{T_1}\left(\om_1(F_{X_2}(x))\right)\nonumber
\\&\leq& \tilde{r}_{T_2}\left(\om_1(F_{X_2}(x))\right).\label{eq3.44}
\end{eqnarray}
\\Case II: Let $T_2$ be DRHR. Then, from (\ref{eq3.43}), we have, for all $x\in[c,d]$,
\begin{eqnarray}
\tilde{r}_{T_2}\left(\om_1(F_{X_2}(x))\right)&\geq&
\tilde{r}_{T_2}\left(\om_1(F_{X_1}(x))\right)\nonumber
\\&\geq& \tilde{r}_{T_1}\left(\om_1(F_{X_1}(x))\right).\label{eq3.45}
\end{eqnarray}
Thus, from (\ref{eq3.44}) and (\ref{eq3.45}), (\ref{eq3.41}) holds.
Now, (\ref{eq3.42}) holds if, and only if, for all $x\in[c,d]$,
$\om'_1(F_{X_2}(x)) f_{X_{2}}(x)\geq \om'_1(F_{X_1}(x))
f_{X_{1}}(x)$, which means that
\begin{eqnarray}F_{X_2}(x)\om'_1(F_{X_2}(x))
\tilde{r}_{X_{2}}(x)\geq F_{X_1}(x)\om'_1(F_{X_1}(x))
\tilde{r}_{X_{1}}(x).\label{eq3.46}
\end{eqnarray}
This holds on using the fact that $x\omega'_1(x)$ is decreasing in
$x\in[0,1)$, and $X_1\leq_{rhr} X_2$.
\begin{remark}\normalfont
 It is to be noted that (i) $\omega_1(x)=-(\ln x)^2$, (ii) $\omega_1(x)=-((1-x)/x)^2$, (iii) $\omega_1(x)=-((1-x)^\alpha /x^{\alpha-1})^2, \alpha>1$ satisfy the condition in Theorem \ref{th3.4}.
$\hfill\Box$\end{remark} \par With the following counterexample, we
show that the condition `$x\omega'_1(x)$ is decreasing in
$x\in[0,1)$' in the above theorem cannot be dropped.
\begin{counterexample}\normalfont
Take $\omega_1(x)=\frac{x}{1-x}$, $x\in[0,1)$ which does not satisfy
the condition in Theorem \ref{th3.4}. Let $X_1$ and $X_2$ follow
exponential distribution with failure rates $\beta_{1}$ and
$\beta_{2}$ respectively. Then $X_{1}\leq_{rhr} X_{2}$ if
$\beta_{1}\geq \beta_{2}$. Let both $T_1$ and $T_2$ follow Weibull
distribution with distribution function
$F_{T_1}(x)=F_{T_2}(x)=1-e^{-x^k}$, $x\geq 0$, $0<k\leq 1$. Here
both $T_1$ and $T_2$ are DRHR. It is easy to verify that, for $\beta_{1}=3$, $\beta_{2}=1.5$, and $k=0.2$,
$F_{T_2}[\om_1(F_{X_2}(x))]/F_{T_1}[\om_1(F_{X_1}(x))]$ is neither
increasing nor decreasing in $x$. Thus, $R_{2}$ does not always
dominate $R_{1}$ with respect to the reversed hazard rate
order.$\hfill\Box$
\end{counterexample}
\par It is well known that although a random variable having support
$[0,\infty)$ cannot be IRHR (increasing in reversed hazard rate) and
most of the well known distributions are DRHR (see Sengupta and
Nanda\citealp{sen}), a distribution function with finite support or
the support of the form $(-\infty,b)$, $0<b<\infty$, can be IRHR
(see Block et al.\citealp{block}). A random variable (or
equivalently its distribution) is said to have IRHR property if
$\tilde{r}(x)$ is increasing in $x$. Below we give some conditions
under which $Z_{2}$ dominates $Z_{1}$ in reversed hazard rate order.
The proof is similar to that of Theorem \ref{th3.4}, and hence
omitted.
\begin{thm}\label{th3.5}
 Let $(1-x)\omega'_1(x)$ be increasing in $x\in[0,1)$. Suppose that the following conditions hold:
 \begin{enumerate}
  \item [(i)] $T_1\leq_{rhr}T_2$, and $T_1$ or $T_2$ is IRHR;
  \item [(ii)] $X_1\geq_{hr} X_2$.
 \end{enumerate}
Then $R_1\leq_{rhr} R_2$.
\end{thm}
\begin{remark}\normalfont
 It is to be noted that (i) $\omega_1(x)=-\ln (1-x)$, (ii)
 $\omega_1(x)=x/(1-x)$, satisfy the condition in Theorem
 \ref{th3.5}.$\hfill\Box$\end{remark}
\begin{remark}\normalfont A random variable $Z$, with finite mean $\mu$, having the following distribution function has IRHR
property.
$$F_{Z}(x)=\left(\frac{bp-\mu+x(1-p)}{b-\mu}\right)^{p/(1-p)},
~x\in(-\infty,b],~p>1.$$
\end{remark}
\section{Conclusion}\label{secCon} In this paper, we study different
reliability properties of Transformed-Transformer family of
distributions, a recently developed method of generating new family
of distributions from some existing distributions. A wide range of
distributions can be generated by using this method (Alzaatreh et
al.\citealp{alz2,alz3}; Lee et al.\citealp{lee}). Several well known
continuous distributions, e.g., beta-generated family of
distributions, Weibull, generalized gamma, exponentiated-Weibull,
gamma-Pareto, exponentiated-exponential distributions, etc. (see
Alzaatreh et al.\citealp{alz2,alz3}) are found to be special cases
of this newly generated distributions. It is shown that these newly
generated distributions are very flexible and are capable of fitting
various types of data (Alzaatreh et al.\citealp{alz3}). The results
studied here will help to decide which distribution to be used,
based on the ageing properties of the newly generated distributions,
and also to find the better one in terms of various stochastic
orders.

\subsection*{Acknowledgements:} Nil Kamal Hazra sincerely acknowledges the financial support from the University of the Free
State, South Africa. The support received from IISER Kolkata to
carry out this research work is gratefully acknowledged by Pradip
Kundu. The financial support from NBHM, Govt. of India (vide Ref.
No. 2/48(25)/2014/NBHM(R.P.)/R\&D II/1393 dt. Feb. 3, 2015) is duly
acknowledged by Asok K. Nanda.


\begin{thebibliography}{00}
\bibitem[(2012)]{aki} Akinsete A, Famoye F, Lee C (2012) The beta-Pareto distribution. Statistics 42(6):547-563.
\bibitem[(2012)]{als} Alshawarbeh E, Lee C, Famoye F (2012) Beta-Cauchy distribution. Journal of Probability and Statistical
Science 10:41-58.
\bibitem[(2013a)]{alz2} Alzaatreh A, Lee C, Famoye F (2013a) A new method for generating families of continuous
distributions. Metron 71(1):63-79.
\bibitem[(2013b)]{alz1} Alzaatreh A, Famoye F, Lee C (2103b) Weibull-Pareto distribution and its applications. Communications in
Statistics-Theory and Methods 42(9):1673-1691.
\bibitem[(2014)]{alz3} Alzaatreh A, Lee C, Famoye F (2014) On generating T-X family of distributions using quantile functions. Journal of Statistical Distributions and
Applications 1(2).
\bibitem[(1985)]{azz} Azzalini A (1985) Class of distributions which includes the normal ones. Scandinavian Journal of Statistics 12:171-178.
\bibitem[(2011)]{bar1} Barreto-Souza W, Cordeiro GM, Simas AB (2011) Some results for beta Fr\'{e}chet distribution. Communications
in Statistics-Theory and Methods 40(5):798-811.
\bibitem[(2010)]{bar2} Barreto-Souza W, Santos AHS, Cordeiro GM (2010) The beta generalized exponential
distribution. Journal of Statistical Computation and Simulation
80(2):159-172.
\bibitem[(1998)]{block} Block HW, Savits TH, Singh H (1998) The reversed hazard
rate function. Probability in the Engineering and Informational
Sciences 12(01):69-90.
\bibitem[(1942)]{bur} Burr IW (1942) Cumulative frequency functions. The Annals of Mathematical Statistics
13:215-232.
\bibitem[(2011)]{cor1} Cordeiro GM, de Castro M (2011) A new family of generalized distributions. Journal of Statistical
Computation and Simulation 81(7):883-898.
\bibitem[(2013)]{cor2} Cordeiro GM, Cristino T, Hashimoto EM, Ortega EM (2013) The beta generalized Rayleigh distribution
with applications to lifetime data. Statistical Papers 54:133-161.
\bibitem[(1969)]{eld} Elderton WP, Johnson NL (1969) Systems of frequency curves. Cambridge University Press, London.
\bibitem[(2002)]{eug} Eugene N, Lee C, Famoye F (2002) Beta-normal distribution
and its applications. Communications in Statistics-Theory and
Methods 31(6):497-512.
\bibitem[(2005)]{fam} Famoye F, Lee C, Olumolade O (2005) The beta-Weibull distribution. Journal of Statistical Theory and
Applications 4(2):121-136.
\bibitem[(1994)]{joh} Johnson NL, Kotz S, Balakrishnan  N (1994) Continuous univariate distributions. Vol. 1, John Wiley and Sons, New
York.
\bibitem[(2009)]{jon} Jones MC (2009) Kumaraswamy's distribution: A beta-type distribution with tractability advantages. Statistical
Methodology 6:70-81.
\bibitem[(2007)]{kon} Kong L, Lee C, Sepanski JH (2007) On the properties of beta-gamma distribution. Journal of Modern Applied Statistical Methods 6(1):187-211.
\bibitem[(2013)]{lee} Lee C, Famoye F, Alzaatreh A (2013) Methods for generating families of univariate continuous distributions
in the recent decades. WIREs Computational Statistics 5(3):219-238.
\bibitem[(2006)]{nad} Nadarajah S, Kotz S (2006) The beta exponential distribution. Reliability Engineering and System Safety
91:689-697.
\bibitem[(2010)]{khob} Khodabin M, Ahmadabadi AR (2010) Some properties of generalized gamma
distribution. Mathematical Sciences 4(1):9-28.
\bibitem[(1974)]{ram1} Ramberg JS, Schmeiser BW (1974) An approximate method for generating asymmetric random
variables. Communications of the Association for Computing Machinery
17:78-82.
\bibitem[(1979)]{ram2} Ramberg JS, Tadikamalla PR, Dudewicz EJ, Mykytka EF (1979) A probability distribution and its uses
in fitting data. Technometrics 21:201-214.
\bibitem[(1999)]{sen} Sengupta D, Nanda AK (1999) Log-concave and concave distributions in
reliability. Naval Research Logistics 46(4):419-433.
\bibitem[(1962)]{stacy} Stacy EW (1962) A generalization of the gamma distribution. The Annals of
Mathematical Statistics 33:1187-1192.

\end{thebibliography}
\end{document}